\documentclass[journal=jctcce,manuscript=article]{achemso}
\setkeys{acs}{articletitle = true}

\usepackage{verbatim}
\usepackage{amsmath,amssymb,amsfonts,amsthm}
\usepackage[latin1,utf8]{inputenc}
\usepackage{graphicx}
\usepackage{booktabs}
\usepackage{subfig}
\usepackage{caption}
\usepackage{float}
\usepackage{color}
\usepackage{mathtools}
\usepackage{multirow}
\usepackage{natbib}
\usepackage[pdftex,bookmarks=false,colorlinks=true,linkcolor=blue,citecolor=blue,filecolor=black,urlcolor=blue]{hyperref}
\usepackage{a4wide,xcolor,eucal,enumerate,mathrsfs}
\usepackage[normalem]{ulem}
\usepackage{amsmath,amssymb,epsfig,amsthm} %,bbm
\usepackage[utf8]{inputenc}
\usepackage{psfrag}          % replace text in figures

\DeclareMathOperator{\Trace}{\mathrm{Tr}}

\DeclareMathOperator{\Epot}{\mathrm{E_{pot}}}

\newcommand{\R}{\mathbb{R}}

\newcommand{\MM}{\mathcal M}

\newcommand{\restr}[1]{\lower3pt\hbox{$|_{#1}$}}

\newcommand{\gammatau}{\gamma^{\tau}}
\newcommand{\dxN}{\mathrm{d}x_1\dots\mathrm{d}x_N}
\newcommand{\dxjN}{\mathrm{d}x_1\dots \hat{\mathrm{d}x_j} \dots \mathrm{d}x_N}
\newcommand{\utau}{u^{\tau}}%\nolimits}
\newcommand{\dens}{\rho}

\numberwithin{equation}{section}

\usepackage{color}
\usepackage{pstricks}
\usepackage{ifthen}
\newrgbcolor{simocolor}{.7 .1 .1}
\newrgbcolor{augucolor}{.1 .1 .7}
\usepackage{csquotes}

\newboolean{DEBUG}
\setboolean{DEBUG}{true}

\ifthenelse {\boolean{DEBUG}}
{}

\ifthenelse {\boolean{DEBUG}}
{}

\graphicspath{{Figures/}}
\title{Kinetic correlation functionals from the entropic regularisation of the strictly-correlated electrons problem}
\author{Augusto Gerolin}
\email{augustogerolin@gmail.com}
\affiliation
{Department of Theoretical Chemistry and Amsterdam Center for Multiscale Modeling, FEW, Vrije Universiteit, De Boelelaan 1083, 1081HV Amsterdam, The Netherlands}
\author{Juri Grossi}
\email{j.grossi@vu.nl}
\affiliation
{Department of Theoretical Chemistry and Amsterdam Center for Multiscale Modeling, FEW, Vrije Universiteit, De Boelelaan 1083, 1081HV Amsterdam, The Netherlands}
\author{Paola Gori-Giorgi}
\email{p.gorigiorgi@vu.nl}
\affiliation
{Department of Theoretical Chemistry and Amsterdam Center for Multiscale Modeling, FEW, Vrije Universiteit, De Boelelaan 1083, 1081HV Amsterdam, The Netherlands}
\DeclareUnicodeCharacter{2212}{-}
\begin{document}
\begin{abstract}
In this work we study the entropic regularisation of the strictly-correlated-electrons formalism, discussing the implications for density functional theory, and establishing a link with earlier works on quantum kinetic energy and classical entropy. We carry out a very preliminary investigation (using simplified models) on the use of the solution of the entropic regularised problem to build approximations for the kinetic correlation functional at large coupling strengths. We also analyze lower and upper bounds to the Hohenberg-Kohn functional using the entropic regularized strictly-correlated-electrons problem. 
\end{abstract}

\section{Introduction}
Despite all their successes, present approximations for the exchange-correlation (XC) functional of Kohn-Sham (KS) density functional theory (DFT) are still plagued by the so-called strong-correlation problem:\cite{MarHea-MP-17} typically, the approximations fail when the physics of the system under study differs too much from the non-interacting one of the KS reference system.

 The leading term of the strong-coupling limit of the DFT adiabatic connection (strictly-correlated electrons - SCE - functional), equivalent to a semiclassical limit ($\hbar\to 0$) at fixed one-electron density, gives access to the exact XC functional in the extreme case that the kinetic energy is neglected with respect to the electron-electron interactions.\cite{ButDepGor-PRA-12,CotFriKlu-CPAM-13, CotFriKlu-ARMA-18,Lew-CRM-18} This strictly-correlated regime is complementary to the one described by the non-interacting KS system. By applying uniform-coordinate scaling, one sees that this limit captures the right physics for low-density systems, i.e., when the average electron-electron distance is much larger than the Bohr radius.\cite{GorSeiVig-PRL-09,MenMalGor-PRB-13} Indeed, when used as an approximation for the XC functional in the self-consistent KS scheme, SCE provides results that get closer and closer to the exact ones as the system is driven to lower and lower density.\cite{MalGor-PRL-12,MalMirCreReiGor-PRB-13,MenMalGor-PRB-14,MirSeiGor-PRL-13} However, with the exception of interesting models for electrons confined at the interface of semiconductor heterostructures,\cite{GhoGucUmrUllBar-NP-06,RonCavBelGol-JCP-06,GhoGucUmrUllBar-PRB-07,MalMirCreReiGor-PRB-13,MenMalGor-PRB-14} chemical systems are never close to this extreme case. Yet, the SCE mathematical structure can be simplified and re-scaled to design functionals for the electron-electron interaction at physical coupling strength,\cite{VucGor-JPCL-17,Vuc-JCTC-19,VucGou-JCP-19} or can be used to build interpolations between the KS and the SCE limits.\cite{SeiPerLev-PRA-99,SeiPerKur-PRL-00,ZhoBahErn-JCP-15,BahZhoErn-JCP-16,VucIroSavTeaGor-JCTC-16,VucIroWagTeaGor-PCCP-17,ZarNeiParPee-PRB-17, FabGorSeiDel-JCTC-16,VucGorDelFab-JPCL-18,GiaGorDelFab-JCP-18,Con-PRB-19, FabSmiGiaDaaDelGraGor-JCTC-19} While these strategies are both very promising, as, for example, they can describe accurately the H$_2$ and H$_2^+$ dissociation curves in the KS spin-restricted formalism,\cite{VucGor-JPCL-17} their main problem is that they do not capture the effects of the kinetic correlation energy, which is known to play a crucial role in the description of strongly-correlated systems in the KS setting,\cite{BuiBaeSni-PRA-89,HelTokRub-JCP-09,TemMarMai-JCTC-09,YinBroLopVarGorLor-PRB-16} with its functional derivative displaying non-intuitive features such as ``peaks'' and ``steps''.\cite{BuiBaeSni-PRA-89,GriLeeBae-JCP-96,HelTokRub-JCP-09,TemMarMai-JCTC-09,HodKraSchGro-JPCL-17,GiaVucGor-JCTC-18} 

The next leading term in the strong-coupling expansion, corresponding to zero-point oscillations in a metric dictated by the density,\cite{GorVigSei-JCTC-09} provides a ``first-order'' kinetic-correlation energy correction,\cite{GroKooGieSeiCohMorGor-JCTC-17} but it is difficult to evaluate in the general case, with its functional derivative displaying features that are too extreme.\cite{GroSeiGorGie-PRA-19} Moreover, this way to do the strong-coupling expansion is not the right one for problems such as bond breaking excitations, because in a molecular system the density close to the atoms remains high: only when we drive the whole system to low density the expansion is really able to capture the right physics.\cite{CorNieLee-PRA-19} 
The purpose of this work is to explore a different route, based on the entropic regularization of Optimal Transport \cite{Gal-CEPR-10,Cut-ANIPS-13, Gal-BOOK-18, PeyCut-FTML-19}, which has been studied in mathematics and economics but also, more recently, has been applied in Data Sciences and Statistical Inference (see, for instance ref~\citenum{PeyCut-FTML-19} and references therein). 

The OT formulation of the SCE functional\cite{ButDepGor-PRA-12,CotFriKlu-CPAM-13} triggered cross fertilization between two different research fields, which led to several formal proofs, setting the SCE limit on firm grounds\cite{CotFriKlu-ARMA-18,Lew-CRM-18,ColDiM-INC-13,ColDepDim-CJM-15}, as well as to new ideas and algorithms.
\cite{Nen-PhD-16,MenLin-PRB-13,FriVog-SIMAJMA-18,KhoLinLin-ARXIV-19,KhoYin-SIAMJSC-19}
Here we focus on the entropic regularization of the SCE problem\cite{Nen-PhD-16, SeiDiMGerNenGieGor-arxiv-17, GerKauRaj-ARXIV-19}, and explore whether this extension can be used to build approximations for the kinetic correlation energy functional and, more generally, to gain new insight in the problem of describing and understanding strong correlation within DFT. As we will explain, the entropic regularization of the SCE problem brings in a new link and perspective on the seminal work of Sears, Parr and Dinur~\cite{SeaParDin-IJC-80} on the relation between various definitions of entropy, information theory, and kinetic energy. Moreover, the formalism is quite general and could also be applied to other interactions and other kind of particles, for example if one wants to treat the nuclei in a quantum DFT setting.\cite{TaoYanHam-JCP-19}.

The paper is organized as follows: in Sec~\ref{sec:entropic} we introduce the theoretical aspects and describe the general form of the solution of the entropic regularization of the SCE functional. In order to illustrate its main properties,  we present simple analytical and numerical examples in Sec~\ref{sec:examples}. We then compare, in Sec~\ref{sec:HK-Entro}, the entropic-regularized SCE functional with the Hohenberg-Kohn functional, discussing inequalities and approximations, with the corresponding numerical and analytical studies in Sec~\ref{sec:comp}. Conclusions and future perspectives are discussed in Sec~\ref{sec:conc}. 

\section{The entropic regularization of the SCE functional}\label{sec:entropic}
Let $\dens(x)$, with $x\in \mathbb{R}^D$, be a density such that $\int_{\mathbb{R}^D}\dens=N$. The SCE functional is defined as
\begin{equation}\label{eq:SCEusual}
	V_{ee}^{\rm SCE}[\dens]=\inf_{\Psi\to\dens}\langle\Psi|V_{ee}|\Psi\rangle,
\end{equation}
i.e., as the infimum, over all possible fermionic wavefunctions having the prescribed density $\dens$, of the expectation value of the electron-electron repulsion operator
\begin{equation}\label{eq:Veedef}
	V_{ee}(x_1,\dots,x_N) = \sum_{1\leq i<j\leq N}v_{ee}(x_i,x_j), \qquad v_{ee}(x,y)=\frac{1}{|x-y|}.
	\end{equation}
We have an infimum in eq~\eqref{eq:SCEusual} because the minimum is attained not on the space of wave functions $\Psi$ (with $\Psi,\nabla\Psi\in L^2({\mathbb{R}^{DN}})$) but on the larger space of probability measures (in physicists/chemists language, by allowing also Dirac-delta distributions)\cite{CotFriKlu-CPAM-13,CotFriKlu-ARMA-18}.  We denote probability measures as $\gamma(x_1,\dots,x_N)$. In a loose way we identify
\begin{equation}\label{def_gamma}
	\gamma(x_1,\dots x_N)= |\Psi(x_1,\dots,x_N)|^2,
\end{equation}
even if $\gamma$ lives in a larger space (i.e., it is allowed to become a distribution). To illustrate what is meant, consider the simple case of $N=2$ and $D=3$. Then the minimizer of eq~\eqref{eq:Veedef} has been proven\cite{ButDepGor-PRA-12,CotFriKlu-CPAM-13} to be always of the SCE form\cite{Sei-PRA-99,SeiGorSav-PRA-07}
\begin{equation}\label{eq:SCEsolutionN2}
\gamma^\mathrm{SCE}(x_1,x_2) =  \frac{1}{2}\dens(x_1)\delta(x_2-f(x_1)),
\end{equation}
which is zero everywhere except on the 3D manifold $x_2=f(x_1)$, parametrized by the co-motion function (or optimal map) $f:\R^3\to\R^3$, with the position of the first electron dictating the position of the second (strict correlation). Notice that the SCE functional has been recently proven to yield the asymptotic low-density (or strong-coupling, or $\hbar\to 0$) limit of the universal Hohenberg-Kohn (HK) functional.\cite{Lew-CRM-18,CotFriKlu-ARMA-18,CotFriKlu-CPAM-13}

On one hand, the fact that the infimum in eq~\eqref{eq:SCEusual} is attained on a probability measure (i.e.,  $\gamma^\mathrm{SCE}$ is concentrated on a low-dimensional manifold of the full configuration space) is exactly what makes the SCE mathematical structure and its density dependence much more accessible than the HK functional.\cite{SeiGorSav-PRA-07,MalGor-PRL-12,MalMirCreReiGor-PRB-13,MenMalGor-PRB-14,VucGor-JPCL-17,BahZhoErn-JCP-16,CorKarLanLee-PRA-17} On the other hand, the challenge of including the effects of kinetic correlation energy stems exactly from the fact that $\gamma^\mathrm{SCE}$ has infinite kinetic energy. We know that in the exact HK functional, even when very close to the SCE limit, kinetic energy will ``spread out'' a little bit the optimal $\gamma$, making it a true $|\Psi|^2$. The zero-point energy (ZPE) expansion gives a recipe for this spreading out, but, as mentioned, in a rather complicated way.\cite{GorVigSei-JCTC-09,GroKooGieSeiCohMorGor-JCTC-17,GroSeiGorGie-PRA-19} Here we consider a particular definition of entropy, used in the OT as a computational regularization, to realize this ``spreading''.

Since it has been proven \cite{Lew-CRM-18,CotFriKlu-ARMA-18} that the fermionic statistics has no effect on the value of $V_{ee}^{\rm SCE}[\dens]$, we work directly in terms of $\gamma(x_1,\cdots,x_N)$, which has the loose sense of eq~\eqref{def_gamma}. 
We then consider the following minimization problem
\begin{equation}\label{eq:EntrFunctional}
F_{\rm entr}^{\tau}[\dens]= \min_{\gamma\to\dens} E^{\tau}[\gamma],
\end{equation}
where the ``entropic'' functional $E^\tau[\gamma]$ is defined for $\tau>0$ as:
\begin{equation}\label{eq:energytau}
 E^{\tau}[\gamma] = V_{ee}[\gamma] -\tau\, S[\gamma]
 \end{equation}
 with
 \begin{align}\label{eq:defVeefunc}
V_{ee}[\gamma] & = \int_{\R^{DN}} V_{ee}(x_1,\dots,x_N)\gamma(x_1,\dots,x_N)\dxN \\
\label{eq:defEntropy}
S[\gamma] & =  -\int_{\R^{DN}} \gamma(x_1,\dots,x_N)\log\gamma(x_1,\dots,x_N)\dxN.
\end{align}
We stress that the entropy term $S:\MM(\R^{DN})\to \R\cup\lbrace +\infty\rbrace$ is defined on the set of signed measures $\MM(\R^{DN})$ such that $\int \gamma = 1$ and it is defined as $S[\gamma] = -\int \gamma\log\gamma$ if $\gamma$ is a probability density and $S[\gamma] = +\infty$ otherwise. 
These conditions force the probability measures to be a probability density $\gammatau$ in $\R^{DN}$ and not a Dirac delta on a manifold as, for example, $\gamma^{\rm{SCE}}$ of eq~\eqref{eq:SCEsolutionN2}, since minus $S[\gamma^{\rm{SCE}}]$ would be equal to $+\infty$. 
The constraint $\gamma\to\dens$ reads explicitly
    \begin{equation}
N\int\gamma(x_1,\dots,x_N)\mathrm{d}x_1\dots\hat{\mathrm{d}x_j}\dots\mathrm{d}x_N=\dens(x_j),\quad\forall j~\in\lbrace1,\dots,N\rbrace,
\end{equation}
where the notation $\hat{\mathrm{d}x}_j$ means that we do not integrate over the variable $x_j$.

We point out that the problem \eqref{eq:EntrFunctional}, typically with $N=2$ and $v_{ee}(x,y)$ in eq~\eqref{eq:Veedef} equal to the $p$-distance $\vert x-y\vert^p$ ($p\geq 1)$, is being studied in different fields, including Probability Theory (e.g. \cite{Leo-DCDSA-14,DimGer-ARXIV-19}), Machine Learning (e.g. \cite{Cut-ANIPS-13, PeyCut-FTML-19}), Scientific computing \cite{BenCarNen-ARXIV-15}, Statistical Physics\cite{KoeDelOrl-PRL-19,KoeDelOrl-PRE-19}, Economics \cite{Gal-BOOK-18}. In the following, we want to analyze the entropic regularization \eqref{eq:energytau} in the framework of the DFT formalism.

First, we remark that the the problem~\eqref{eq:EntrFunctional} admits a unique solution $\gammatau$, since the functional $E^{\tau}[\gamma]$ is strictly convex in $\gamma$. Second, we can fully characterise the unique solution in~\eqref{eq:EntrFunctional}: as shown, for instance in refs~\citenum{BorLewNus-JFA-94, Leo-DCDSA-14}, and \citenum{DimGer-ARXIV-19}, $\gammatau$ is the solution of~\eqref{eq:EntrFunctional} if, and only if,
\begin{equation}\label{eq:defai}
\gammatau(x_1,\dots,x_N) = \prod_{i=1}^N a^\tau(x_i) e^{-V_{ee}(x_1,\dots,x_N)/\tau}, 
\end{equation}
where $a^\tau(x):\R^D\to\R$ is the so-called \textit{entropic weight} and is fixed by the density constraint 
\begin{equation}\label{eq:aj}
a^\tau(x_j) \int_{\R^{D(N-1)}} \prod_{i\neq j} a^\tau(x_i) e^{-\frac{V_{ee}(x_1,\cdots,x_N)}{\tau}}\dxjN = \frac{\dens(x_j)}{N}, \quad \forall \, j = 1,\dots, N.
\end{equation}
The entropic weight $a^\tau(x)$ can be written as an exponential of the entropic one-body potential $u^\tau(x)$,
\begin{equation}\label{eq:entropicpot}
a^\tau(x) = e^{\frac{u^\tau(x)}{\tau}},
\end{equation}
 with $u^\tau(x)$ having the usual physical interpretation of DFT, as (minus) the potential that enforces the density constraint. The theorems behind eqs~\eqref{eq:defai}-\eqref{eq:entropicpot} are non-trivial, and we point to ref~\citenum{DimGer-ARXIV-19} for a rigorous proof in the case of bounded interactions $v_{ee}$, and to the Appendix for more details on how this potential appears as the dual variable with respect to the density, as in standard DFT. Here, in order to provide an intuitive idea of the role of the entropic weight, we consider the problem \eqref{eq:EntrFunctional} in a box $[-L,+L]^{DN} \subset \R^{DN}$ and we minimize $E^{\tau}[\gamma]$ with respect to $\gamma$ without fixing the density constraint, obtaining the usual result, i.e. that $\gammatau$ is a Gibbs state
\begin{equation}
    \gamma=Z e^{-\frac{V_{ee}(x_1,\cdots,x_N)}{\tau}}, \quad \text{where} \quad Z = N\left(\int_{\R^{DN}} e^{-\frac{V_{ee}}{\tau}} \dxN\right)^{-1}.
\end{equation} 
This clearly shows that the entropic weight $a^\tau(x)=e^{\frac{u^\tau(x)}{\tau}}$ is a Lagrange multiplier to enforce the constraint $\gamma\to\rho$ in \eqref{eq:EntrFunctional}. The solution $\gammatau$ in \eqref{eq:defai} can then be written as
 \begin{equation}\label{eq:gammawithu}
 \gammatau(x_1,\dots,x_N) = \exp\left(\frac{\sum_{i=1}^N u^\tau(x_i)-V_{ee}(x_1,\dots,x_N)}{\tau}\right).
 \end{equation}
We should remark at this point that the one-body potential $u^\tau(x)$ is not gauged to approach zero when $|x|\to\infty$ but it is shifted by a constant $C^\tau[\dens]$,
\begin{equation}
	u^\tau(|x|\to\infty)=C^\tau[\dens].
\end{equation}
When $\tau\to 0$ this constant ensures that
\begin{equation}
V_{ee}(x_1,\dots,x_N)-\sum_{i=1}^N u^0(x_i)\ge 0.
\end{equation}
This way, we see that $\gamma^{\tau\to 0}$ of eq~\eqref{eq:gammawithu} becomes more and more concentrated on the manifold where $V_{ee}(x_1,\dots,x_N)-\sum_{i=1}^N u^0(x_i)$ is minimum and equal to 0. We can interpret $V_{ee}(x_1,\dots,x_N)-\sum_{i=1}^N u^0(x_i)$ as an hamiltonian without kinetic energy whose minimising wavefunction is constrained to yield the given density $\rho$ by the one-body potential $u^0(x)$. In fact, this is the hamiltonian that appears as leading term in the strong-coupling limit of the usual density-fixed DFT adiabatic connection,\cite{LanPer-SSC-75,SeiGorSav-PRA-07} whose minimising $\gamma$ (if we relax the space in which we search for the minimum) will be zero everywhere except on the manifold where $V_{ee}(x_1,\dots,x_N)-\sum_{i=1}^N u^0(x_i)$ has its global minimum. This is exactly the SCE manifold parametrised by the co-motion functions.

Notice that the constant $C^0[\dens]=\lim_{\tau\to 0} C^\tau[\dens]$ is precisely the same,\cite{VucLevGor-JCP-17} in the strong-coupling limit of DFT, as the one discussed by Levy and Zahariev in the context of KS DFT.\cite{LevZah-PRL-14} In fact, since the potential $u^0(x)$ is gauged at infinity to a constant that guarantees that the minimum of $V_{ee}(x_1,\dots,x_N)-\sum_{i=1}^N u^0(x_i)$ is equal to zero, and since the optimal $\gamma^{\tau\to 0}$ will be concentrated on the manifold where the minimum is attained, by simply taking the expectation value of $V_{ee}(x_1,\dots,x_N)-\sum_{i=1}^N u^0(x_i)$ on $\gamma^{\tau\to 0}$ we obtain
\begin{equation}\label{eq:SCEasintofu}
	V_{ee}^{\rm SCE}[\rho]=\int_{\R^D} \rho(x) u^0(x) \rm{d}x.
\end{equation}
Moreover, we also have that $u^0$ is a functional derivative with respect to $\rho$ (gauged to a constant at infinity) of $V_{ee}^{\rm SCE}[\rho]$.\cite{MalGor-PRL-12,MalMirCreReiGor-PRB-13} If we use $V_{ee}^{\rm SCE}[\rho]$ as an approximation for the Hartree and exchange-correlation energy, as in the KS SCE approach,\cite{MalGor-PRL-12,MalMirCreReiGor-PRB-13,MenMalGor-PRB-14} then eq~\eqref{eq:SCEasintofu} is exactly the condition imposed by Levy and Zahariev\cite{LevZah-PRL-14} to their constant shift.

\subsection{Interpretation of the parameter $\tau$ and of the entropy $S[\gamma]$}\label{subsec:interpr}
 One can simply regard $\tau>0$ as a parameter interpolating between two opposite regimes: the strictly-correlated one and the uncorrelated bosonic case with the prescribed density. 

\begin{figure}
\centering
\subfloat[][$\tau=0.1$]
   {\includegraphics[width=.33\textwidth]{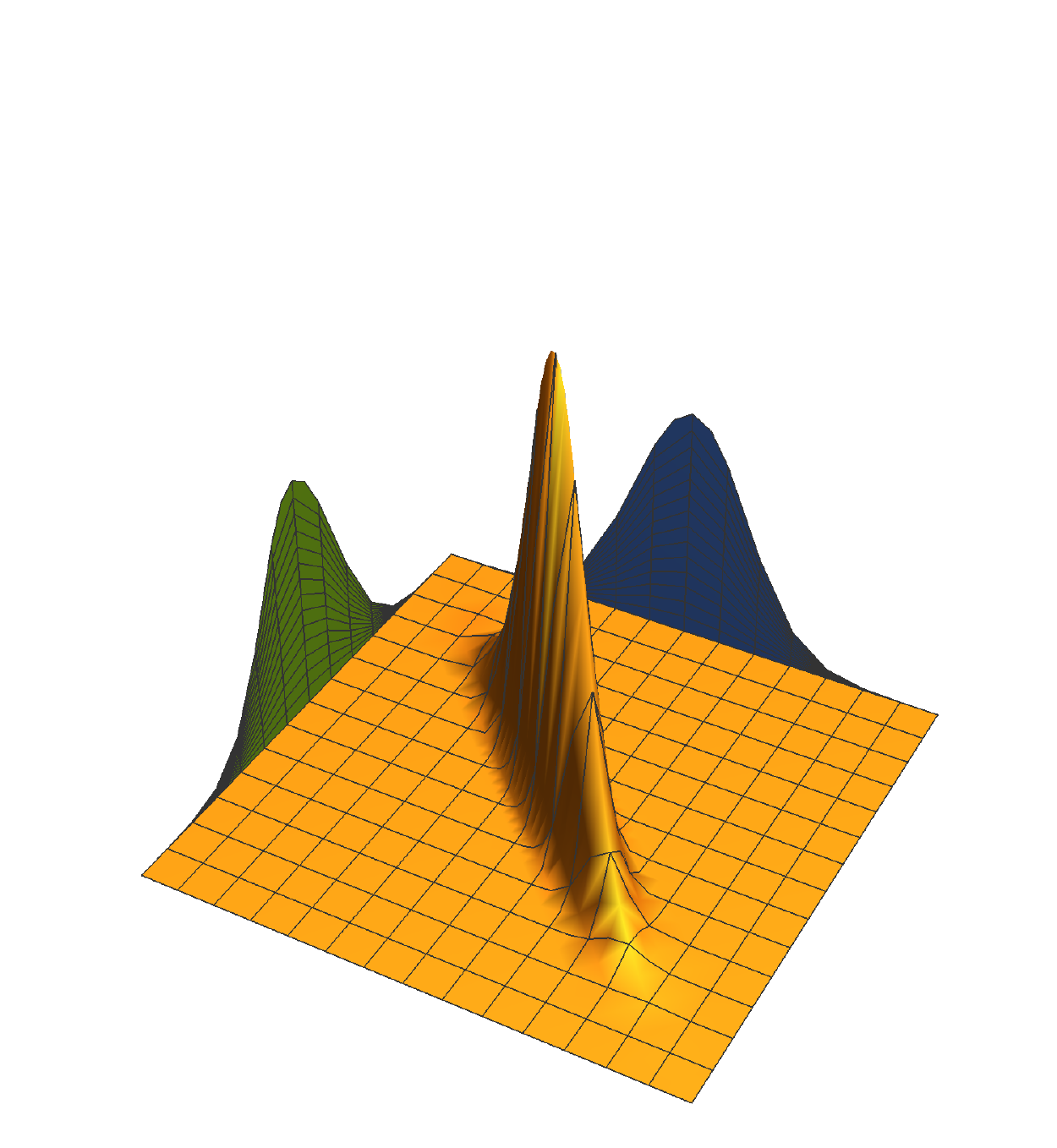}} 
\subfloat[][$\tau=1$]
   {\includegraphics[width=.33\textwidth]{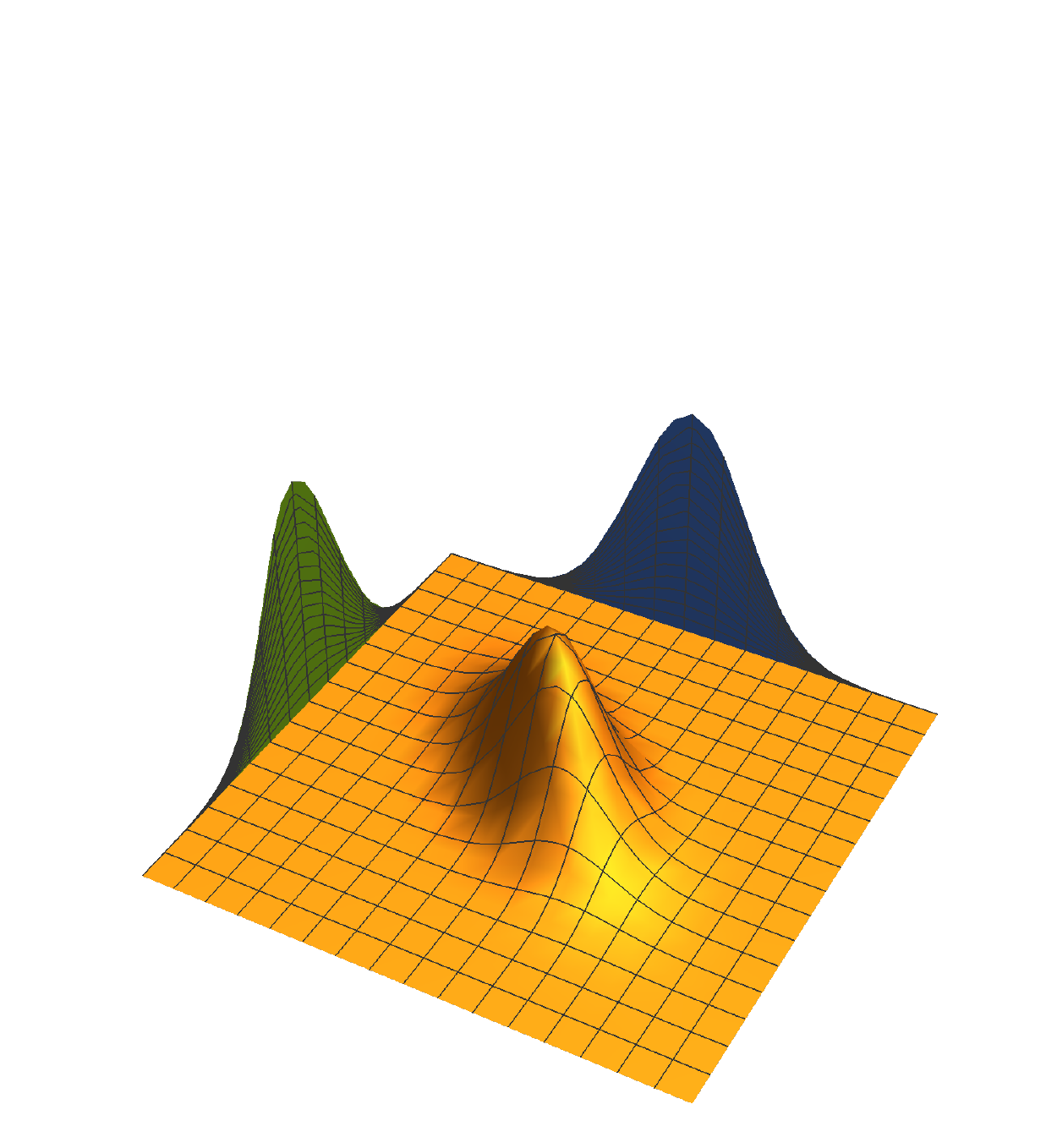}} 
\subfloat[][$\tau=5$]
   {\includegraphics[width=.33\textwidth]{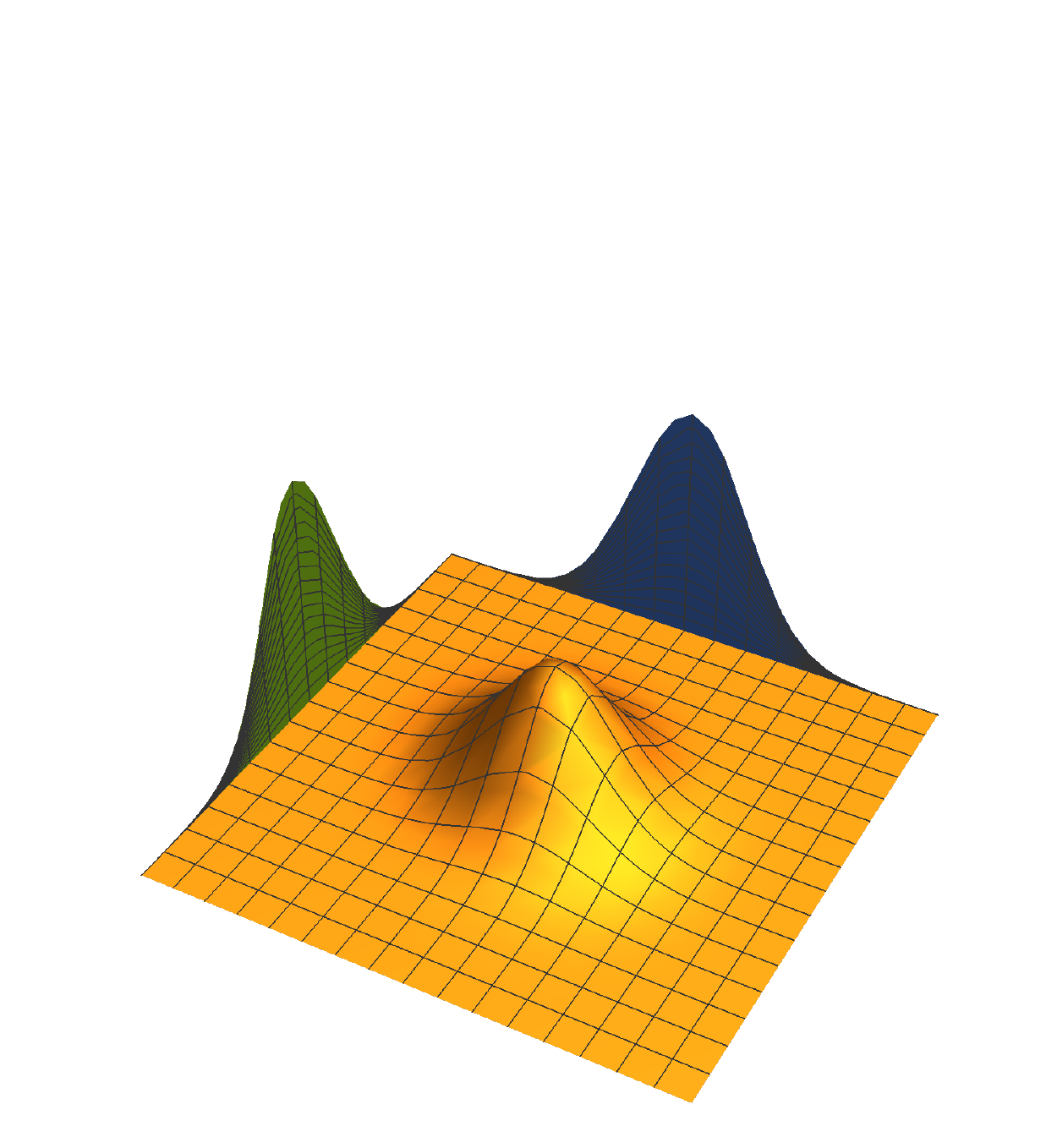}} 
\caption{The optimal $\gamma(x_1,x_2)$ for the interaction $v_{ee}(x,y)=-(x-y)^2$, at different values of $\tau$. Notice that the marginals $\frac{\dens(x_1)}{2}$, $\frac{\dens(x_2)}{2}$ remain the same at all $\tau$, while $\gamma$ evolves from the strictly correlated regime $\gamma^{\tau=0.1} \sim \gamma^{0}(x_1,x_2) = \frac{\rho(x_1)}{2}\delta(x_2-f(x_1))$, with $f(x) = -x$, to the symmetric uncorrelated one $\gamma^{\tau=5} \sim \gamma^{\infty}= \frac{\dens(x_1)\dens(x_2)}{4}$. See section~\ref{sec:Harm} for a fully analytical description of this example.}
\label{fig:gammaharm}
\end{figure}

In fact, when $\tau\to 0$ the problem~\eqref{eq:EntrFunctional} becomes the one defined by the SCE functional of eq~\eqref{eq:SCEusual} \cite{GerKauRaj-ARXIV-19}, and, as just discussed, $\gamma^\tau$, given by eq~\eqref{eq:gammawithu},
in this limit is more and more concentrated on the manifold on which $V_{ee}(x_1,\dots,x_N)-\sum_{i=1}^N u^0(x_i)=0$. In the case $N=2$, this is exactly the three-dimensional manifold $\{x_1=x,x_2=f(x)\}$ parametrised by the co-motion function (or optimal map) $f(x)$ of eq~\eqref{eq:SCEsolutionN2}. To visualise this, in fig.~\ref{fig:gammaharm}, we show a simple example with $N=2$ particles in 1D, having a gaussian density,  whose interaction is repulsive harmonic. In panel (a) of this figure we show $\gamma^{\tau\to 0}(x_1,x_2)$, which is concentrated on the manifold $x_2=f(x_1)$, where for this special case $f(x)=-x$. For $N>2$, we usually (but not always) also have a three-dimensional manifold parametrised by cyclical maps $f_i(x)$.\cite{SeiGorSav-PRA-07,SeiDiMGerNenGieGor-arxiv-17}

When $\tau\to\infty$, the problem~\eqref{eq:EntrFunctional} converges to the one of maximizing $S[\gamma]$ alone under the constraint $\gamma\to\dens$, 
\begin{equation}
\lim_{\tau\to\infty}\tau^{-1}F_{\rm entr}^\tau[\dens] = \min_{\gamma\to \dens}\lbrace -S[\gamma]  \rbrace=\max_{\gamma\to \dens}\lbrace S[\gamma]  \rbrace.
\end{equation}
This is equivalent to maximize the entropy of $\gamma$ relative to the product state $\prod^N_{i=1}\frac{\dens(x_i)}{N}$. In fact, with $\tilde{\dens}(x)=\dens(x)/N$,
\begin{align}\label{eq:pippone}
S[\gamma]&=-\int\gamma(x_1,\dots,x_N)\log(\gamma(x_1,\dots,x_N))\mathrm{d}x_1\dots\mathrm{d}x_N\nonumber\\&=
-\int\gamma(x_1,\dots,x_N)\log\bigg(\frac{\gamma(x_1,\dots,x_N)}{\prod_i\tilde{\dens}(x_i)}\prod_i\tilde{\dens}(x_i)\bigg)\mathrm{d}x_1\dots\mathrm{d}x_N\nonumber\\&=
-\int\gamma(x_1,\dots,x_N)\log\bigg(\frac{\gamma(x_1,\dots,x_N)}{\prod_i\tilde{\dens}(x_i)}\bigg)-\int\gamma(x_1,\dots,x_N)\log\bigg(\prod_i\tilde{\dens}(x_i)\bigg)\nonumber\\&=
-\int\gamma(x_1,\dots,x_N)\log\bigg(\frac{\gamma(x_1,\dots,x_N)}{\prod_i\tilde{\dens}(x_i)}\bigg)-\sum_i\int\gamma(x_1,\dots,x_N)\log(\tilde{\dens}(x_i))\nonumber\\&=
-\int\gamma(x_1,\dots,x_N)\log\bigg(\frac{\gamma(x_1,\dots,x_N)}{\prod_i\tilde{\dens}(x_i)}\bigg)-N\int\tilde{\dens}(x_1)\log(\tilde{\dens}(x_1))\mathrm{d}x_1.
\end{align}
Since the density is held fixed, the second term in the last line is a constant during the maximization. Gibbs inequality applied to the relative entropy (first term in the last line) then gives $S[\gamma]\leq S[\prod^N_{i=1}\frac{\dens(x_i)}{N}]$, and the optimal $\gamma$ that maximizes $S[\gamma]$ is then the uncorrelated product state. Equation \eqref{eq:pippone} also shows that the entropy $S[\gamma]$ with fixed one-electron density is a relative entropy (Kullback–Leibler divergence) with respect to the uncorrelated product, a.k.a. non-interacting bosonic state with the prescribed density. In other words, at fixed density the uncorrelated product is the probability density whose support has the maximal volume. This is illustrated, again in the simple 1D case with repulsive harmonic interactions, in panel (c) of fig~\ref{fig:gammaharm}, where we also show, in panel (b) a case in between these two extremes.
 
The problem~\eqref{eq:energytau} has been already used as an auxiliary functional to compute numerically the solutions of~\eqref{eq:SCEusual}. In fact, the entropy term reinforces the uniqueness of the minimiser in~\eqref{eq:energytau}. The parameter $\tau$ in this case regularizes the problem of~\eqref{eq:SCEusual} (``spreading out'' the support of $\gamma$, as in fig~\ref{fig:gammaharm}), which can be solved via the Sinkhorn algorithm.\cite{Cut-ANIPS-13,BenCarNen-ARXIV-15}

We should emphasize that, as eq~\eqref{eq:pippone} clearly shows, the entropy $S[\gamma]$ used here is different from the quantum mechanical entropy of finite-temperature DFT (see \citenum{Mer-PR-65, PriGraBur-PRL-16,BurSmiGraPri-PRB-16}, \citenum{PriPitGro-BOOK-14}  and references therein), which is defined in terms of density matrices and favors mixed states. Here $S[\gamma]$ can be interpreted in terms of mutual information (or discrimination information), measuring how a probability $\gamma$ differs from a reference distribution, in this case the uncorrelated product. 
A related definition and interpretation in terms of the Kullback–Leibler divergence, including its link to kinetic energy, was considered by Sears, Parr and Dinur~\cite{SeaParDin-IJC-80} in the context of DFT. The link between various definitions of entropy and kinetic energy is also present in several works in the literature; in particular the link with the kinetic correlation energy is conjectured in ref~\citenum{Del-IJQC-15}.

Before comparing the functional $F_{\rm entr}^{\tau}[\dens]$ with the Hohenberg-Kohn functional close to the strong-coupling regime, we find it important to illustrate the formalism just introduced with simple examples.

\section{Analytic and numerical examples of the entropic regularization problem}\label{sec:examples}

\subsection{Harmonic interactions case}\label{sec:Harm}
We start by considering the repulsive and attractive harmonic interaction $v_{ee}(x,y)=\xi(x-y)^2$, with $\xi=\pm 1$. This interaction is interesting not only because it allows for analytic solutions with which one can fully illustrate the formalism, but also because it arises as leading term in the effective interaction between electrons bound on two different distant neutral fragments (dispersion). In fact, if we keep the densities of the two fragments frozen at their isolated ground-state values (a variational constraint that has several computational advantages and can lead to very accurate or even exact results\cite{KooGor-JPCL-19}), minimizing the dipolar interaction, which contains terms like $x_1 x_2$ orthogonal to the bond axis and $-z_1 z_2$ parallel to it, is equivalent to minimizing the repulsive and attractive harmonic interaction, respectively. This is simply because $\pm x_1 x_2$ differs from $\mp \frac{1}{2}(x_1-x_2)^2$ only by one-body terms, which do not affect the minimizer when the density is held fixed.
Another case in which harmonic interactions could be interesting is if we want to treat (some) nuclei quantum mechanically.
\subsubsection*{(a) $N=2$}
To allow for a completely analytic solution we  fix the one-body density to be a Gaussian. This is exactly the Drude quantum oscillator model for the coarse-grained dispersion between two fragments \cite{Bad-JCP-57,FerDiSAmbCarTka-PRL-15} when we forbid the oscillator density to change with respect to its isolated value (a constraint that gives the exact result for the dispersion coefficient $C_6$ between two oscillators, exactly like in the case of the H atom\cite{KooGor-JPCL-19}). Since the dipolar interaction separates in the 3 spatial directions, we can consider the one-dimensional case with
\begin{equation}
\dens(x)=\frac{2}{\sqrt{\pi}\sigma}e^{-\frac{x^2}{\sigma^2}}.
\end{equation}
In the following we use the notation $x=x_1$ and $y=x_2$ for the coordinates of the two particles in 1D. By writing $\gamma^\tau(x,y)=a^\tau(x)a^\tau(y)e^{\frac{-v_{ee}(x,y)}{\tau}}$ and dividing both sides of eq~\eqref{eq:aj} by $a^\tau(x)$, we see that eq~\eqref{eq:aj} becomes, after writing $a^\tau(x)=e^{\frac{u^\tau(x)}{\tau}}$,
\begin{equation}\label{eq:ex1}
\int_{-\infty}^{+\infty}e^{\frac{u^\tau(x)-v_{ee}(x,y)}{\tau}}\mathrm{d}x=\frac{2}{\sqrt{\pi}\sigma}e^{-\frac{y^2}{\sigma^2}}e^{-\frac{u^\tau(y)}{\tau}}.
\end{equation}
As previously discussed, if we find the explicit form for $u^{\tau}(x)$ that satisfy Eq.~\eqref{eq:ex1}, then we have found the optimal one.
We then first assume that the solution $u^\tau$ can be restricted to a class of 2nd degree polynomials 
\begin{equation}
u^{\tau}(x)=a_\tau x^2+c_\tau,
\end{equation}
and verify that indeed it is possible to obtain a solution of this kind, which amounts 
to solve the system of equations
\begin{equation}
\begin{cases}
\frac{a_\tau^2-2 a_\tau \xi }{\tau  (a_\tau-\xi )}=-\frac{1}{\sigma ^2}\\\frac{2 a_\tau -2  \xi }{\tau  (a_\tau-\xi )}c_\tau+\frac{1}{2} \log
   \left(-\frac{\pi  \tau }{a_\tau+1}\right)=-\log \left(\sqrt{\pi } \sigma
   \right)\\
\end{cases}
\end{equation} which yields, choosing the negative solution
\begin{equation}
\begin{cases}
a_{\tau }=-\frac{\sqrt{4 \xi ^2 \sigma ^4+\tau ^2}-2 \xi  \sigma ^2+\tau }{2
   \sigma ^2}\\
c_{\tau }=-\frac{1}{4} \tau \log \left(\frac{2 \tau \sigma^4\pi^2 }{\sqrt{4 \xi ^2
   \sigma ^4+\tau ^2}-2 (\xi +1) \sigma ^2+\tau }\right).
\end{cases}
\end{equation}
Defining $l^2=\sqrt{4 \sigma ^4+\tau ^2}+\tau$, the corresponding minimizing $\gamma^\tau(x,y)$ reads
\begin{equation}
\gamma^\tau(x,y)=\frac{\sqrt{\frac{l^2-2 (\xi +1) \sigma ^2}{\tau }}}{\sqrt{2} \pi  \sigma ^2}e^{-\frac{l^2 \left(x^2+y^2\right)}{2 \tau \sigma
   ^2}+\frac{2 \xi  x
   y}{\tau }},
\end{equation} 
and it is shown at different values of $\tau$ (with $\sigma=1$ and $\xi=-1$) in fig.~\ref{fig:gammaharm}, where, as anticipated in sec~\ref{subsec:interpr}, we see the transition from the SCE-like state at small $\tau$, to the uncorrelated product state at large $\tau$.

\subsubsection*{(b) $N>2$, $D=1$}
In the case when $N>2$, the first equation of the system in.~\ref{eq:ex1} reads
\begin{equation}
\underbrace{\int_{-\infty}^{+\infty}e^{\frac{\sum_{i=2}^N u^\tau(x_i)-\sum_{i>j\geq 2}v_{ee}(x_i,x_j)}{\tau}}\mathrm{d}x_2\dots\mathrm{d}x_N}_{= I(x_1)}=\dens(x_1)e^{-\frac{u^\tau(x_1)}{\tau}}.
\end{equation}
with
\begin{equation}
I(x_1)=\frac{\pi ^{\frac{N-1}{2}} \tau ^{\frac{N-1}{2}} }{\sqrt{\left(a_{\tau }+\xi \right)
   \left(a_{\tau }+\xi  N\right){}^{N-2}}}\exp \left(\frac{(N-1) \left(4 c_{\tau } \left(a_{\tau }+\xi \right)+4 \xi  x_1^2 a_{\tau }\right)}{4 \tau  \left(a_{\tau }+\xi \right)}\right).
\end{equation}
By arguing similarly as in the in the previous paragraph, one can obtain that the solution of the equation
\begin{equation}
\log(I(x_1))=-\frac{\utau(x_1)}{\tau}+\log(\dens(x_1))
\end{equation} is given by 
\begin{equation}
\begin{cases}
a_\tau&=-\frac{\sqrt{\left(\xi  N \sigma ^2+\tau \right)^2-4 \xi  \sigma ^2 \tau }+\xi  N \sigma
   ^2+\tau }{2 \sigma ^2}\\
c_\tau&=-\frac{\tau }{2 N} \log \left[\frac{(-1)^{N+1} (2 \pi \tau \sigma^2)^{N}
   \left(\sqrt{\left(\xi  N \sigma ^2+\tau \right)^2-4 \xi  \sigma ^2 \tau }-\xi  N
   \sigma ^2+\tau \right)^{2-N}}{8\tau\left(\sqrt{\left(\xi  N \sigma ^2+\tau \right)^2-4 \xi 
   \sigma ^2 \tau }+\xi  (N-2) \sigma ^2+\tau \right)}\right]
\end{cases}
\end{equation}

\subsection{Regularized Coulomb interaction case}\label{sec:regcoul}
\begin{figure}
\centering
\subfloat[][$\tau=0.001$]
   {\includegraphics[width=.31\textwidth]{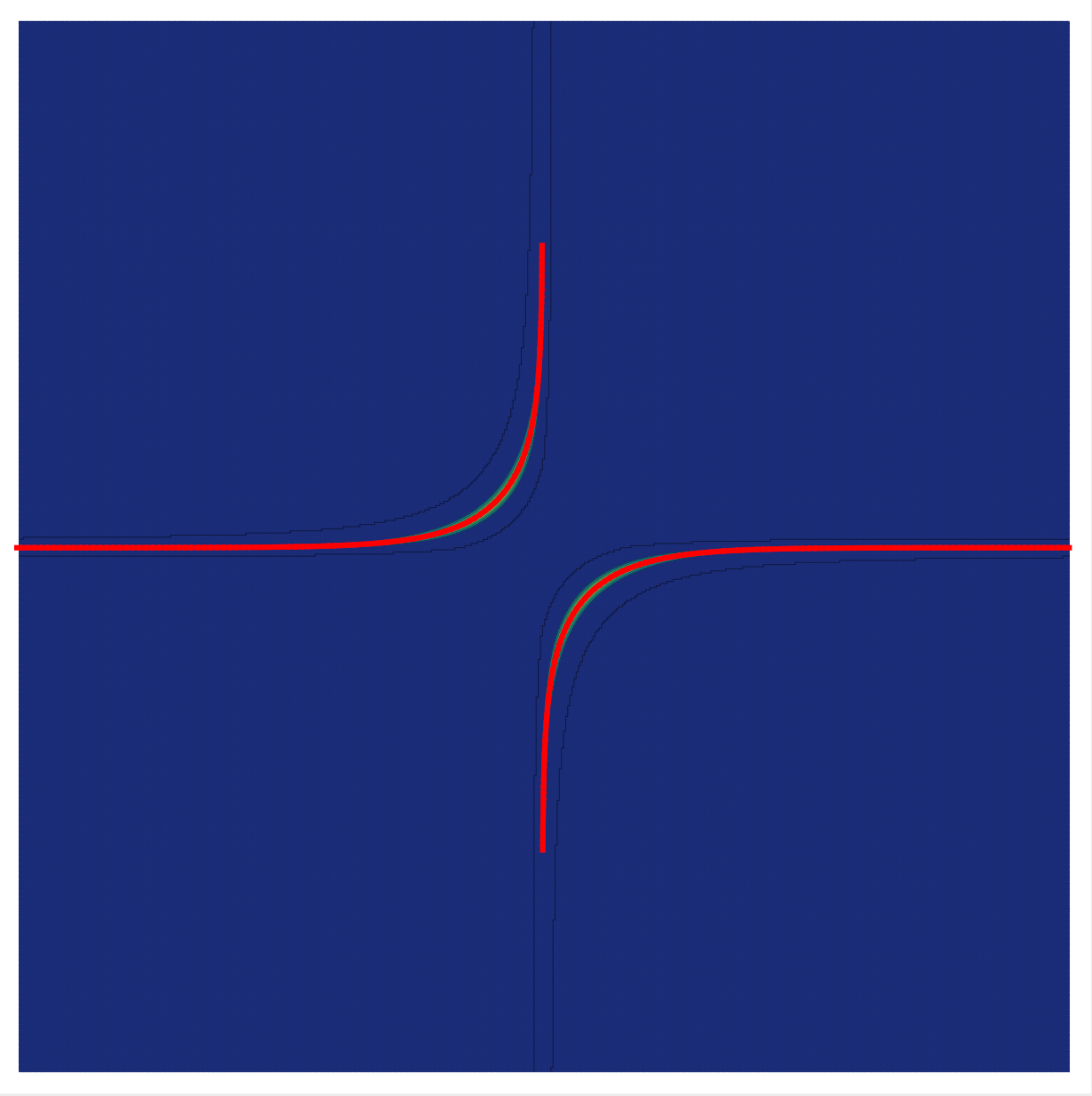}}~~
\subfloat[][$\tau=0.01$]
   {\includegraphics[width=.31\textwidth]{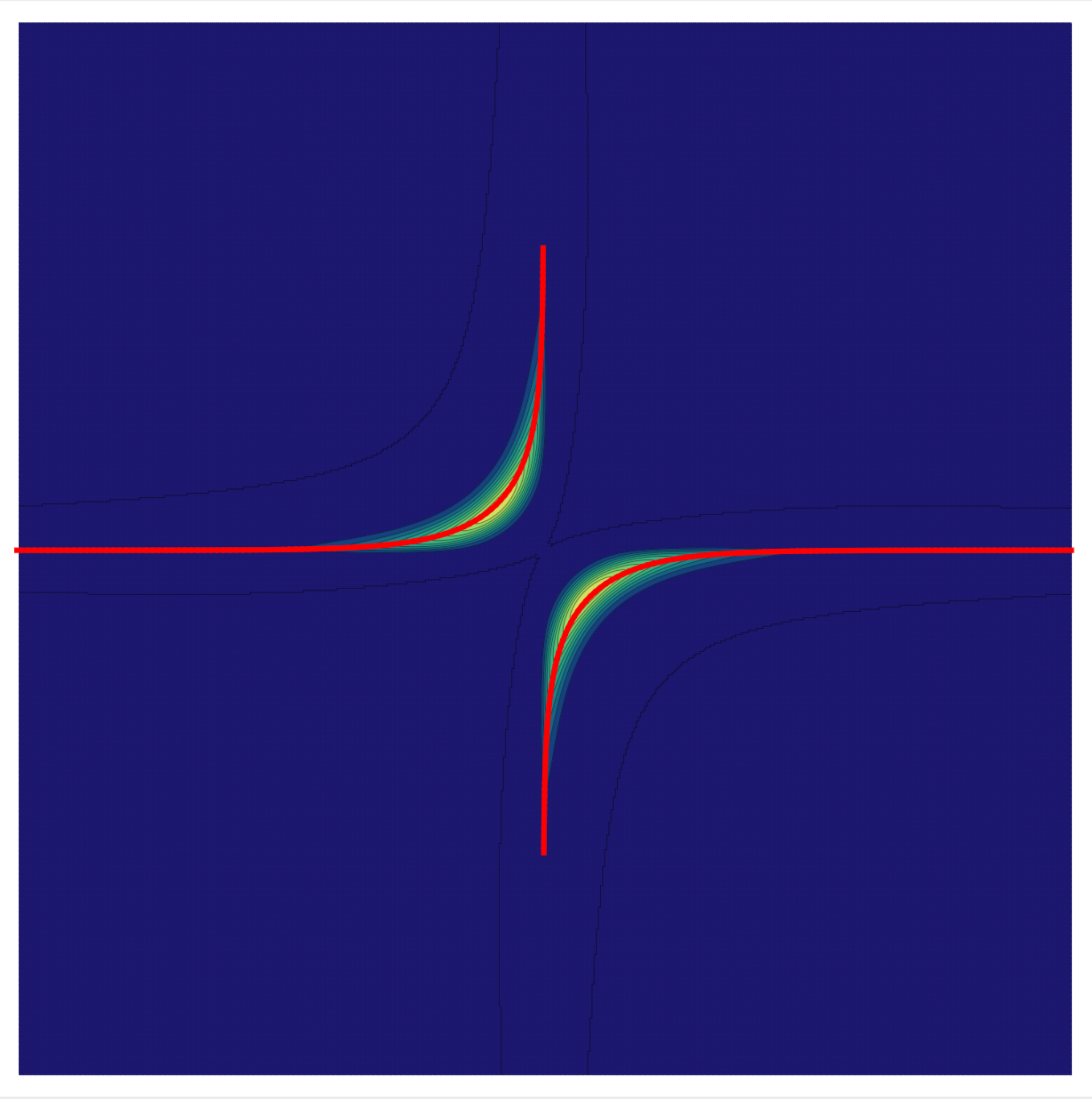}}~~
\subfloat[][$\tau=1$]
   {\includegraphics[width=.31\textwidth]{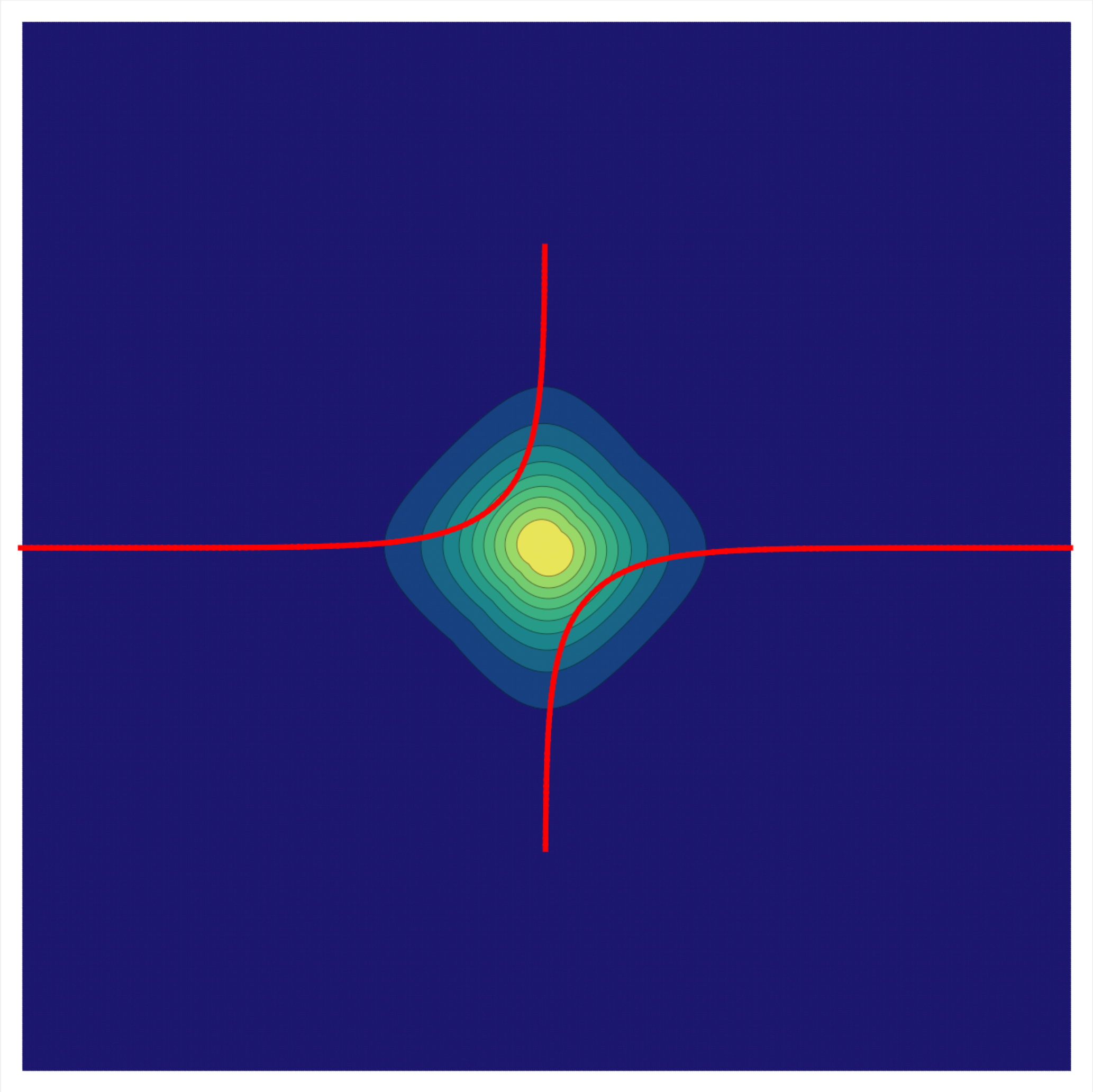}} 
\caption{The support of the optimal $\gamma^{\tau}(x_1,x_2)$ for the interaction $v_{ee}(x,y)=1.07 e^{-\frac{\vert x_1-x_2\vert}{2.39}}$, at different values of $\tau$ for the density \eqref{eq:density}. The red line represents the co-motion function $x_2=f(x_1)$.}
\label{fig:gammacoul}
\end{figure}

For illustrative purposes, we now consider a  1D problem with $N=2$ electrons interacting via the effective Coulomb repulsion, $v_{ee}(x_1,x_2)=1.07 e^{-\frac{\vert x_1-x_2\vert}{2.39}}$, which has been shown in ref~\citenum{BakStoMilWagBurSte-PRB-15} to yield results that closely mimic the 3D electronic structure. In sec~\ref{sec:Coulcomp} we will also consider another 1D interaction, with long-range Coulomb tail, finding results qualitatively very similar. We fix the density to be
\begin{equation}\label{eq:density}
\dens_\mathrm{C}(x)=\mathcal{N}\frac{1}{\cosh(x)},\quad x\in[-10,10].
\end{equation}
The reason to choose this particular density is that it has an exponential decay at large $x$ (similar to an atomic density) and allows for an analytic solution in the the SCE case.\cite{GroKooGieSeiCohMorGor-JCTC-17} 
For the entropic regularization case, however, the solution of the system of equations~\eqref{eq:aj} cannot be obtained analytically, and therefore we have computed it numerically via the Sinkhorn algorithm \cite{Cut-ANIPS-13} (POT library \cite{FlaCou-misc-17}). 
In fig.~\ref{fig:gammacoul} we report our results for the support of $\gamma^\tau$, as $\tau$ increases: in panel (a), corresponding to a small value of $\tau$, we see that $\gamma^\tau(x_1,x_2)$ is different from zero only very close to the manifold parametrized by the co-motion function, $x_2=f(x_1)$, thus becoming a very good approximation for $\gamma^{SCE} = \frac{\rho(x_1)}{2}\delta(x_2-f(x_1))$. We also show, as a tiny red line, the co-motion function $f(x)$ computed analytically\cite{GroKooGieSeiCohMorGor-JCTC-17} from the SCE theory. 
Panel (c) corresponds to a relatively high value of $\tau$, and we see that $\gamma^\tau$ is approaching the uncorrelated bosonic product $\gamma^{\infty}(x_1,x_2) = \frac{\rho(x_1)\rho(x_2)}{4}$, losing any resemblance with the SCE state. The central panel (b) is for us the most interesting: the system is still close to the SCE state, but it has a significant ``spreading'', which could be used to approximate the quantum system close to (but not at) the SCE limit, mimicking the role of kinetic energy. We will explore this possibility in the next two sections.

\section{Comparison with the Hohenberg-Kohn functional}\label{sec:HK-Entro}
In this section we compare the entropic functional $F_{\rm entr}^{\tau}[\dens]$ of eq~\eqref{eq:EntrFunctional} with the Hohenberg-Kohn\cite{HohKoh-PR-64} functional (HK) in its extension of Levy and Lieb\cite{Lev-PNAS-79,Lie-IJQC-83} as a constrained minimisation problem, generalized to arbitrary coupling strength $\lambda\ge 0$,
\begin{equation}\label{eq:Flambda}
	F_\lambda[\dens]= \min_{\Psi\to\dens} \left\{T[\Psi]+\lambda\,V_{ee}[\vert\Psi\vert^2]\right\}
\end{equation}
with
\begin{equation}
	T[\Psi]  = \dfrac{N}{2}\int_{\R^{DN}}\vert \nabla_{1} \Psi(x_1,\dots,x_N)\vert^2 \mathrm{d}x_1\dots \mathrm{d}x_N,
\end{equation}
and $V_{ee}[\gamma]$ defined in eq~\eqref{eq:defVeefunc}. Notice that $F_0[\dens]$ is the Kohn-Sham functional and $F_1[\dens]$ is the Hohenberg-Kohn functional at physical coupling strength.
In particular, we are interested in exploring how the large-$\lambda$ expansion of $\lambda^{-1}F_\lambda[\dens]$ compares with the entropic functional $F_{\rm entr}^{\tau}[\dens]$ at small $\tau$. We already know that the two limits are equal,
\begin{equation}
	\lim_{\tau\to 0} F_{\mathrm{entr}}^{\tau}[\dens]=\lim_{\lambda\to\infty}\frac{F_\lambda[\dens]}{\lambda}=V_{ee}^{\mathrm{SCE}}[\dens],
\end{equation}
but we want to compare how they behave when approaching the SCE limit, slightly ``spreading out'' the optimal $\gamma$ into a $|\Psi|^2$ around the SCE manifold as in panel (b) of fig~\ref{fig:gammacoul}. To begin with, we briefly recall how $F_\lambda[\dens]$ behaves at large $\lambda$, namely\cite{SeiGorSav-PRA-07,GorVigSei-JCTC-09,GroKooGieSeiCohMorGor-JCTC-17}
\begin{equation}\label{eq:subleadingHK}
F_{\lambda}[\dens]\sim\lambda\, V_{ee}^{\mathrm{SCE}}[\dens]+\sqrt{\lambda}\,F^{\mathrm{ZPE}}[\dens],\quad\lambda\to \infty,
\end{equation}
where $F^{\mathrm{ZPE}}[\dens]$ is the zero-point energy functional. Similarly to the functional $S[\gamma]$, the zero point oscillations performed by the $N$ particles around the manifold parametrized by the co-motion functions (optimal maps) $f_i(x)$ allow for the corresponding probability density $\gamma^{\mathrm{ZPE}}$ to provide a finite kinetic energy. Calling $\mathbb{H}(x)$ the Hessian matrix of the function $V_{ee}(x_1,\dots,x_N)-\sum_{i=1}^N u^0(x_i)$ evaluated on the manifold $\{x_1=x,x_2=f_2(x),\dots,x_N=f_N(x)\}$, the two functionals in eq~\eqref{eq:subleadingHK} can be written explicitly as\cite{GorVigSei-JCTC-09}
\begin{align}
F^{\mathrm{ZPE}}[\dens]&=\frac{1}{2}\int_{\mathbb{R}^D}\mathrm{d}x\,\frac{\dens(x)}{N}\,\Trace\left(\sqrt{\mathbb{H}(x)}\right)\\
V_{ee}^{\mathrm{SCE}}[\dens]&=\frac{1}{2}\sum_{i=2}^{N}\int_{\mathbb{R}^D}\mathrm{d}x\,\dens(x)\,v_{ee}(\vert x-f_i(x)\vert).
\end{align}
In particular, due to the virial theorem we can write the $\lambda$-dependent expectation value of the electron-electron interaction and of the kinetic energy operator at large $\lambda$:
\begin{equation}
\begin{cases}
V_{ee}[|\Psi_\lambda[\dens]|^2]&\sim V_{ee}^{\mathrm{SCE}}[\dens]+\frac{F^{\mathrm{ZPE}}[\dens]}{2\sqrt{\lambda}}\\
T[\Psi_\lambda[\dens]]&\sim\sqrt{\lambda}~\frac{F^{\mathrm{ZPE}}[\dens]}{2}
\end{cases},\quad\lambda\to \infty
\end{equation}
where $\Psi_\lambda[\dens]$ is the minimizer of \eqref{eq:Flambda}. We should stress that, while for the leading term in eq~\eqref{eq:subleadingHK} there are rigorous mathematical proofs\cite{Lew-CRM-18,CotFriKlu-ARMA-18}, the term of order $\sqrt{\lambda}$ is a very plausible conjecture,\cite{GorVigSei-JCTC-09} which has been confirmed numerically in some simple cases.\cite{GroKooGieSeiCohMorGor-JCTC-17}

\subsection{Inequalities and approximations}\label{sec:ineq}
First of all, as shown in  ref~\citenum{SeiDiMGerNenGieGor-arxiv-17}, as a simple
consequence of the logarithmic Sobolev inequality for the Lebesgue measure \cite{GozLeo-PTRF-07}, it holds
\begin{equation}
F_{\mathrm{entr}}^{\tau}[\dens]\leq  \frac{F_{\lambda}[\rho]}{\lambda}  \quad \text{ with } \tau = \frac{\pi}{2\lambda} \label{eq:SobolevInequality}
\end{equation}
However, this entropic lower bound to the HK functional can be very loose, as we will show in figs.~\ref{fig:companalitico}-\ref{fig:compnumerico} with some numerical examples. 
We also have
\begin{equation}\label{eq:ineqsuSCE}
V^{\mathrm{SCE}}_{ee}[\dens]\leq F_{\mathrm{entr}}^{\tau}[\dens] -\tau N\int\frac{\dens}{N}\log\left(\frac{\dens}{N}\right)\quad\forall\tau\ge 0,
\end{equation}
simply because this way we have added a positive quantity to the value of $V_{ee}[\gamma]$ obtained with the $\gamma^\tau$ that minimizes eq~\eqref{eq:EntrFunctional}.

A route we explore in this work is the use of the $\gamma^\tau[\rho]$ from an entropic calculation at finite $\tau$ to compute an approximate many-body kinetic energy in the $\lambda\to\infty$ limit,
\begin{equation}
	T_{\mathrm{entr}}^\tau[\dens] = T[\sqrt{\gamma^{\tau}[\rho]}]= \frac{N}{2}\int_{\mathbb{R}^{DN}}\vert \nabla_1 \sqrt{\gamma^\tau(x_1,\dots,x_N)}\vert^2 \mathrm{d}x_1\dots \mathrm{d}x_N,
\end{equation}	
where $\gamma^\tau[\rho]$ is the optimum in the problem \eqref{eq:EntrFunctional} with the given $\dens$.  Since $\gamma^\tau$ has the explicit form \eqref{eq:gammawithu} (in terms of the entropic potential $u^\tau(x)$ that needs to be computed numerically), we obtain
\begin{equation}\label{eq:Tentrexplicit}
	T_{\mathrm{entr}}^\tau[\dens]=\frac{1}{\tau^2}\frac{N}{8}\int \gamma^\tau(x_1,\dots,x_N)\big|\nabla u^\tau(x_1)-\sum_{i=2}^N \nabla v_{ee}(x_1-x_i)\big|^2 \mathrm{d}x_1\dots \mathrm{d}x_N.
\end{equation}
Obviously, $\gamma^\tau$ will not have the right nodal surface and will miss the fermionic character. However, the fermionic statistics is expected\cite{GorVigSei-JCTC-09,GorSeiVig-PRL-09} to appear in $F_\lambda[\dens]$ at large $\lambda$ only through orders $\sim e^{-\sqrt{\lambda}}$, a conjecture that was supported with numerical evidence.\cite{GroKooGieSeiCohMorGor-JCTC-17} The idea is to use the large-$\lambda$ functional as an approximation for the Hartree-exchange-correlation functional, so that the fermionic character will be captured by the KS kinetic energy, similarly to the KS SCE scheme.\cite{MalGor-PRL-12,MalMirCreReiGor-PRB-13,MenMalGor-PRB-14,KhoLinLin-ARXIV-19}
More generally, we will analyze the functional $G_{\lambda}^\tau[\rho]$ defined as
\begin{equation}\label{eq:Gtaudef}
G_{\lambda}^\tau[\rho] = T[\sqrt{\gamma^{\tau}[\rho]}] + \lambda V_{ee}[\gamma^{\tau}[\rho]],
\end{equation}
with $\gamma^{\tau}[\dens]$ the minimizer of $F_{\mathrm{entr}}^\tau[\dens]$. As a consequence of the variational principle, we have for the special case of a $N=2$ closed-shell system
\begin{equation}\label{eq:taulambdainequality}
F_\lambda[\dens]\leq G_{\lambda}^\tau[\rho] ,\quad\forall\lambda,\tau \quad (N=2).
\end{equation}
However, for $N>2$ the inequality will not be valid in general, as $\sqrt{\gamma^{\tau}[\rho]}$ does not have the right fermionic antisymmetry. We still expect it to hold for large $\lambda$ with $\tau\propto \lambda^{-1/2}$, where the energetic difference between fermionic and bosonic minimisers should become exponentially small,\cite{GroKooGieSeiCohMorGor-JCTC-17} of orders $\sim e^{-\sqrt{\lambda}}$.
In the following Sec~\ref{sec:comp} we provide a first explorative study into different ways to find an optimal relation between $\tau$ and $\lambda$, in order to make $G_{\lambda}^\tau[\dens]$ as close as possible to $F_\lambda[\dens]$. Notice that by looking at eq~\eqref{eq:Tentrexplicit} one may expect that $T_{\mathrm{entr}}^\tau[\dens]$ diverges as $1/\tau^2$ for small $\tau$. However, the divergence is milder, because when $\tau\to 0$ the integrand in eq~\eqref{eq:Tentrexplicit} tends to zero, as $\gamma^{\tau\to 0}$ is more and more concentrated on the manifold where $V_{ee}(x_1,\dots,x_N)-\sum_{i=1}^N u^0(x_i)$ is minimum (and stationary, i.e., where its gradient, contained in the modulus square inside the integrand, is zero). 
We believe that $T_{\mathrm{entr}}^\tau[\dens]$ diverges only as $1/\tau$ for small $\tau$, implying that $\tau$ should be proportional to $\lambda^{-1/2}$ to match the large-$\lambda$ expansion of the HK functional, a conjecture that seems to be confirmed by our analytical and numerical results in the next Sec~\ref{sec:comp}. However, we have no rigorous proof for this statement.

\section{Analytical and numerical investigation}\label{sec:comp}
In section~\ref{sec:ineq} a specific relation between $\tau$ and $\lambda$ was used to establish a rigorous inequality, namely eq~\eqref{eq:SobolevInequality}, which holds $\forall\lambda$ when $\tau(\lambda)=\frac{\pi}{2\lambda}$. 
The question we want to address here is whether for a given $\lambda$ (and in particular for large $\lambda$), the inequality \eqref{eq:taulambdainequality} can be sharpened into an equality by tuning $\tau$ according to a general dependence $\tau(\lambda)$. We thus look for $\tau$ that solves
\begin{equation}\label{eq:taulambdaequality}
F_\lambda[\dens]=G_{\lambda}^{\tau(\lambda)}[\rho].
\end{equation}
Although this equation can probably be always solved, at least for large $\lambda$, the real question is whether we can find a reasonably accurate general approximation for the relation between $\tau$ and $\lambda$, as, obviously, we do not want to compute the exact HK functional each time to determine the proper $\tau(\lambda)$. Here we make a very preliminary numerical and analytic exploration, which supports the already conjectured relation $\tau(\lambda)\sim \lambda^{-1/2}$ at large $\lambda$. Finding an approximate $\tau(\lambda)$ that is generally valid, however, remains for now an open challenge, which requires further investigations.

\subsection{Repulsive Harmonic interaction} \label{subsec:harmcomparisons}
\begin{figure}
\centering
\includegraphics[width=1\textwidth]{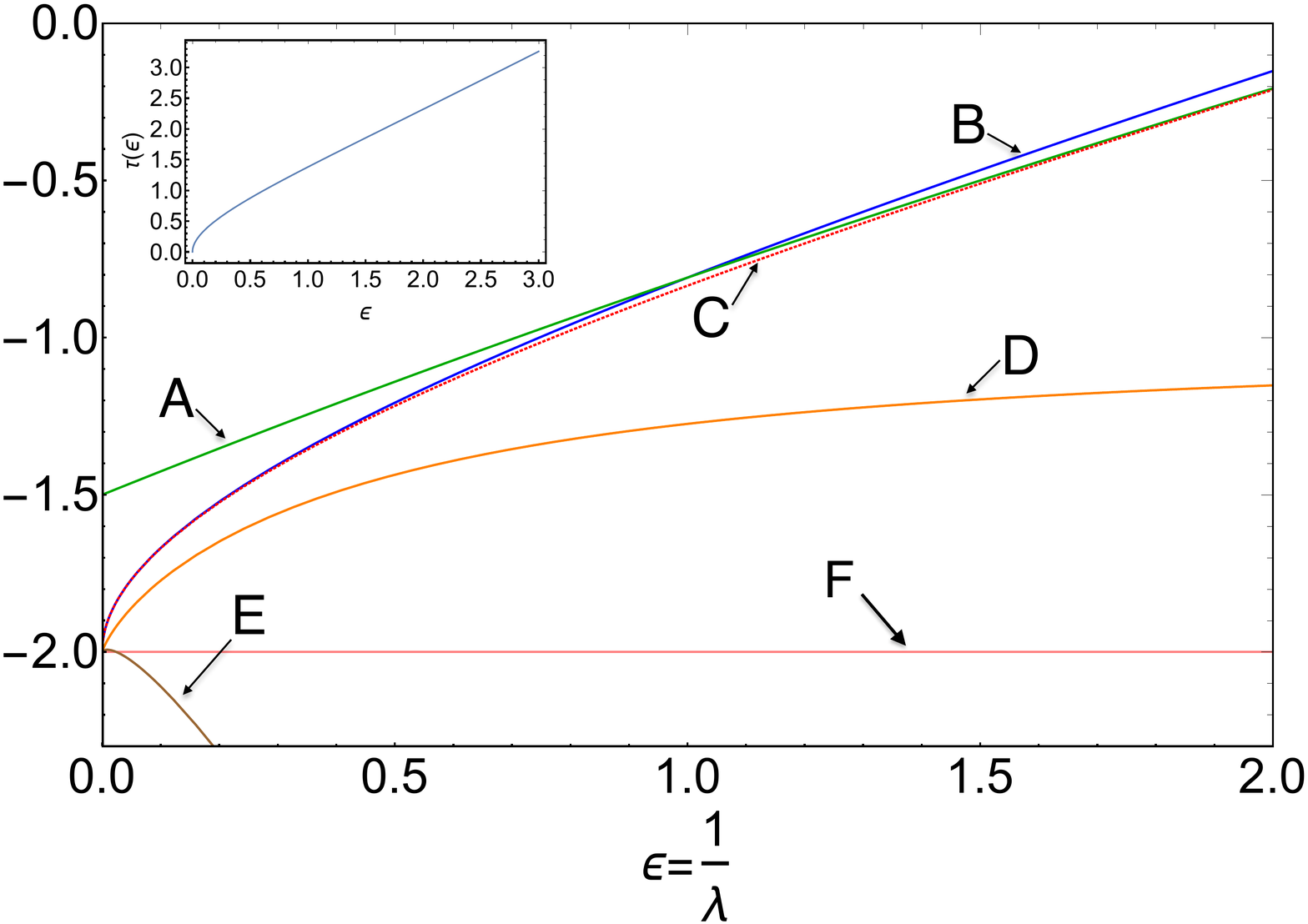}
\caption{The exact expansion of the solution to Eq.~\eqref{eq:taulambdaequality} for the repulsive harmonic interaction and a gaussian density as a function of $\epsilon=\lambda^{-1}$. $A=\epsilon~ G_{1/\epsilon}^{\tau=\epsilon}[\dens]$, $B=\epsilon~ G_{1/\epsilon}^{\tau=\sqrt{\epsilon}}[\dens]$, $C=\epsilon~ F_{1/\epsilon}[\dens]$, $D=F_{\mathrm{entr}}^{(\pi/2)\epsilon}[\dens]-\frac{\pi}{2}\epsilon\int\dens\log\frac{\dens}{2}$, $E=F_{\mathrm{entr}}^{(\pi/2)\epsilon}[\dens]$, $F=V_{ee}^{\mathrm{SCE}}$. See text in sec~\ref{subsec:harmcomparisons} for a detailed explanation.}
\label{fig:companalitico}
\end{figure}

Equation~\eqref{eq:taulambdaequality} can be solved explicitly for the example discussed in section \ref{sec:Harm}, where $N=2$, the density is a Gaussian and the electron-electron interaction is repulsive harmonic. In fact, we start by noticing that the exact wavefunction minimizing $F_{\lambda}[\dens]$ with repulsive harmonic electron-electron interaction and a Gaussian density has the form (see, e.g., the appendix of ref~\citenum{SeiGorSav-PRA-07})
\begin{equation}
\gamma_{\mathrm{exact}}^\lambda(x_1,x_2)=\mathcal{N}_\lambda e^{-\mathcal{C}_\lambda(x_1+x_2)^2-\mathcal{D}_{\lambda}(x_1-x_2)^2},
\end{equation}while
\begin{equation}\label{eq:entrschematico}
\gamma^\tau(x_1,x_2)=\tilde{\mathcal{N}}_\tau e^{-\mathcal{A}_\tau(x_1^2+x_2^2)-\mathcal{B}_{\tau}(x_1-x_2)^2},
\end{equation}implying that $\gamma^\tau$ can always be mapped to $\gamma_{\mathrm{exact}}^\lambda$ by setting 
\begin{equation}
\begin{cases}
2\mathcal{C}_{\lambda}&=\mathcal{A}_\tau\\
2\mathcal{D}_{\lambda}&=\mathcal{A}_\tau-2\mathcal{B}_\tau\\
\end{cases}.
\end{equation}
This implies that, being $\gamma^\tau$ essentially of the exact form for this specific case, we can just evaluate the functional $\tilde{G}_{\lambda}[\gamma^\tau]=T[\sqrt{\gamma^\tau}]+\lambda V_{ee}[\gamma^\tau]$ and minimize it with respect to the coefficients $\mathcal{A}_\tau$,  $\mathcal{B}_\tau$. The constraint $\gamma^\tau\to\dens$ implies
\begin{equation}
\mathcal{A}_\tau=\frac{1}{2} \left(-\sqrt{4 \mathcal{B}_\tau^2+1}+2 \mathcal{B}_\tau+1\right).
\end{equation}
Equation~\eqref{eq:Gtaudef} reads then
\begin{equation}
\tilde{G}_{\lambda}[\gamma^\tau]=\frac{1}{4} \left(\sqrt{4 \mathcal{B}_\tau^2+1}+1\right) -\lambda\frac{\sqrt{4 \mathcal{B}_\tau^2+1}+2 \mathcal{B}_\tau-1}{2 \mathcal{B}_\tau},
\end{equation}
and we obtain the optimal $\mathcal{B}_\tau$ as a function of $\lambda$ by setting
\begin{equation}
\frac{\mathrm{d}\tilde{G}_{\lambda}[\gamma^\tau]}{\mathrm{d}\mathcal{B}_\tau}=0
\end{equation}The only positive solution, $\mathcal{B}_\tau(\lambda)$, provides the answer. In fact, direct comparison of \eqref{eq:entrschematico} with Eq.~\eqref{eq:gammawithu} shows that
\begin{equation}
\mathcal{B}_\tau(\lambda)=\frac{1}{\tau}\Rightarrow\tau(\lambda)=\mathcal{B}_\tau(\lambda)^{-1},
\end{equation}or
%\begin{subequations}
%\begin{eqnarray}
%\tau(\lambda)&=&\frac{2 \sqrt[3]{6} \sqrt[6]{\Delta } \sqrt[4]{\mathcal{J}}}{\sqrt[3]{\lambda } \sqrt{12 \sqrt{\Delta } \sqrt[3]{\lambda }-\sqrt[3]{2} \Delta ^{2/3} \sqrt{\mathcal{J}}+8
 %  \sqrt[3]{3} \sqrt{\mathcal{J}} \lambda ^{2/3}}-\mathcal{J}^{3/4}},\label{eq:univoverotau}\\
 %  \mathcal{J}&= &\sqrt[3]{2} (\Delta  \lambda )^{2/3}-8 \sqrt[3]{3} \lambda ^{4/3},\\ 
 %  \Delta&=&\sqrt{768 \lambda ^2+81}+9 .
%\end{eqnarray}
%\end{subequations}
\begin{subequations}
\begin{eqnarray}
	\tau(\lambda) & = & 2\times 6^{1/3}\Delta^{1/6}\Big(\lambda^{1/3}\Big(8\times
	3^{1/3}\lambda^{2/3}-2^{1/3}\Delta^{2/3}+\frac{12\lambda^{1/3}\sqrt{\Delta}}{\sqrt{2^{1/3}\Delta^{2/3}\lambda^{2/3}-8\times 3^{1/3}\lambda^{4/3}}} \Big)^{1/2} \nonumber \\
	& &-\sqrt{2^{1/3}\Delta^{2/3}\lambda^{2/3}-8\times 3^{1/3}\lambda^{4/3}}\Big)^{-1}\label{eq:univoverotau} \\
	\Delta & = & \sqrt{768 \lambda ^2+81}+9
\end{eqnarray}
\end{subequations}
Eq.~\eqref{eq:univoverotau} has the following asymptotic expansions:
\begin{equation}\label{eq:taulambdaapprox}
\tau(\lambda)\sim\begin{cases}
\frac{1}{\lambda }+\lambda+\mathcal{O}\left(\lambda^3\right),&\quad\lambda\to 0 \\
\sqrt{\frac{1}{\lambda }}+\frac{1}{4 \lambda }+\frac{3}{32\,\lambda^{3/2}}+\mathcal{O}\left(\lambda^{-2}\right),&\quad\lambda\to \infty
\end{cases}
\end{equation}
confirming $\tau(\lambda) \sim \lambda^{-1/2}$ for $\lambda\to\infty$, as discussed in Sec~\ref{sec:ineq}. Both series at small and large $\lambda$ will have a finite radius of convergence since the function $\tau(\lambda)$, eq~\eqref{eq:univoverotau}, has several branch cuts. The exact $\tau(\lambda)$ of eq~\eqref{eq:univoverotau} can be very accurately represented with the following simple Pad\'e approximant that interpolates between the two limits of eq~\eqref{eq:taulambdaapprox}, 
\begin{equation}\label{eq:tauPade}
	\tau_{\rm Pad}(\lambda)=\frac{32 \lambda ^{3/2}+32 \lambda ^2+24 \lambda +15 \sqrt{\lambda }+32}{\lambda(32 \lambda
   ^{3/2}+24 \lambda +15 \sqrt{\lambda }+32)}
\end{equation}
In fig~\ref{fig:companalitico} we compare, as a function of $\epsilon=\lambda^{-1}$, the exact HK functional $\epsilon F_{1/\epsilon}[\rho]$ (curve labelled ``C'') with the results obtained from the functional $G_{\lambda}^\tau[\rho]$ of eq~\eqref{eq:Gtaudef} by using for $\tau(\lambda)$ different approximations. In the curve labelled ``A'' we have used the $\lambda\to 0$ leading term of eq.~\eqref{eq:taulambdaapprox}, $\tau(\lambda)=\lambda^{-1}$ and in the curve labelled ``B'' we have used the $\lambda\to\infty$ leading term, $\tau(\lambda)=\lambda^{-1/2}$. We  see that, this way, we approximate $F_{\lambda}[\dens]$ at different correlation regimes. We also show in the same figure the left-hand side of the inequality~\eqref{eq:ineqsuSCE} when we set $\tau(\lambda)=\frac{\pi}{2\lambda}$, which was found in the inequality \eqref{eq:SobolevInequality}, curve labelled ``D''. As it should, this curve stays above the value of $V_{ee}^{\rm SCE}[\dens]$ (horizontal line, labelled ``F''), but, in this case, it also stays below the HK functional, which is a nice feature, although probably peculiar to the harmonic interaction (see next Sec~\ref{sec:Coulcomp}). We also show the right-hand side of the inequality \eqref{eq:SobolevInequality} (curve labelled ``E''), which, as anticipated, is a very loose lower bound. The result obtained by using the Pad\'e approximant $\tau_{\rm Pad}(\lambda)$ of eq~\eqref{eq:tauPade} in $G_{\lambda}^\tau[\rho]$ is, on the scale of fig~\ref{fig:companalitico}, indistinguishable from the exact curve.

\subsection{Effective Coulomb interaction} \label{sec:Coulcomp}
For an interaction that mimics the electron-electron repulsion in quasi-1D systems there is no analytical computation available. As anticipated in section \ref{sec:regcoul}, we resort to the Sinkhorn algorithm to obtain the quantities of interest, and repeat the computation just done for the harmonic cost. We tested two different interaction forms for $v_{ee}$, namely a regularized Coulomb interaction, and the exponential interaction already used in section \ref{sec:regcoul} to compute $\gamma^\tau$ at various regimes:
\begin{equation}\label{eq:twointeractions}
\begin{cases}
v_{ee}^{\mathrm{reg}}(x)&=\frac{1}{1+\vert x\vert}\\
v_{ee}^{\mathrm{exp}}(x)&=1.07 ~e^{-\frac{\vert x\vert}{2.39}},\\
\end{cases}
\end{equation}
with the same density of eq~\eqref{eq:density}.
\begin{figure}
\centering
\subfloat[][Regularized Coulomb interaction.]
   {\includegraphics[width=.8\textwidth]{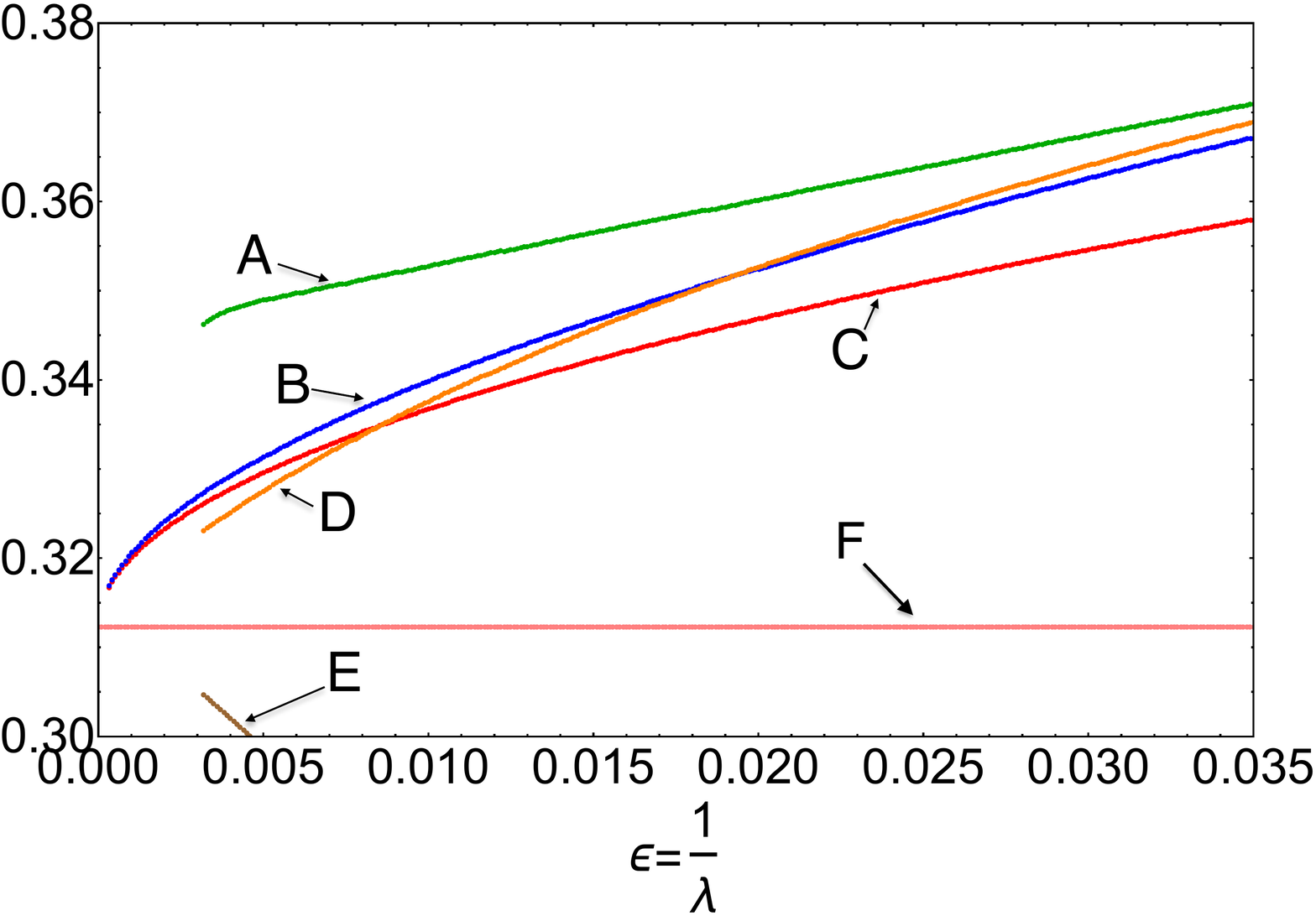}}\\
\subfloat[][Exponential interaction.]
   {\includegraphics[width=.8\textwidth]{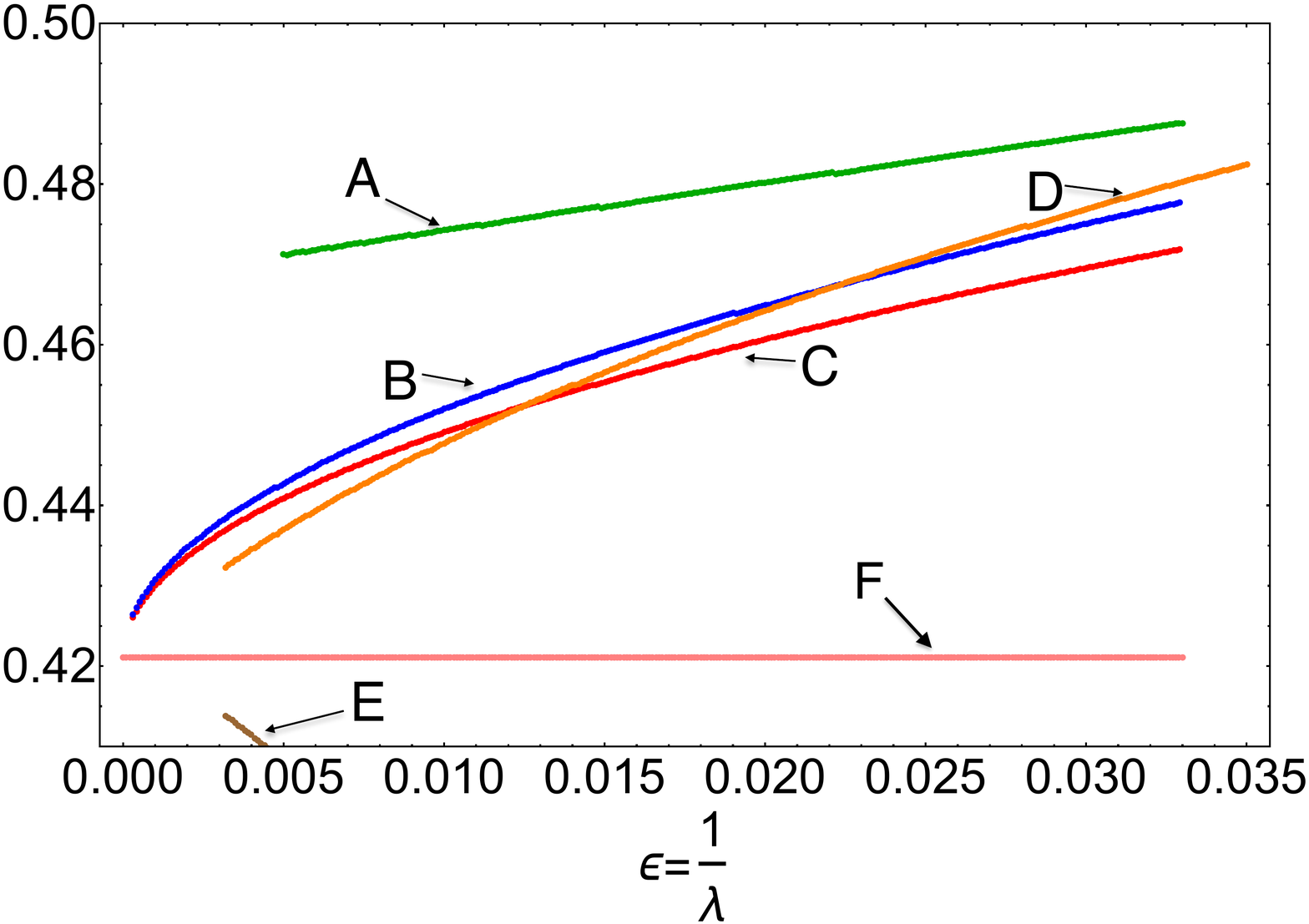}}
\caption{The exact expansion of the solution to Eq.~\eqref{eq:taulambdaequality} for different interactions and density $\dens_C$. $A=\epsilon~ G_{1/\epsilon}^{\tau=\epsilon}[\dens]$, $B=\epsilon~ G_{1/\epsilon}^{\tau=a\sqrt{\epsilon}}[\dens]$, $C=V_{ee}^{\mathrm{SCE}}[\dens]+\epsilon ~F^{\mathrm{ZPE}}[\dens]$, $D=F_{\mathrm{entr}}^{(\pi/2)\epsilon}[\dens]-\frac{\pi}{2}\epsilon\int\dens\log\frac{\dens}{2}$, $E=F_{\mathrm{entr}}^{(\pi/2)\epsilon}[\dens]$. The horizontal line represents $V_{ee}^{\mathrm{SCE}}[\dens]$. The numerical value of $a$ in Eq.~\eqref{eq:taususpect} reads respectively $a=0.27$ (upper figure) and $a=0.32$ (lower figure). See text in sec~\ref{sec:Coulcomp} for a detailed explanation.}
\label{fig:compnumerico}
\end{figure}
In fig~\ref{fig:compnumerico} we  compare $G_{\lambda}^{\tau(\lambda)}[\dens]$, using  different approximations for $\tau(\lambda)$,with the expansion  $\lambda V_{ee}^{\mathrm{SCE}}+\sqrt{\lambda}F^{\mathrm{ZPE}}[\rho]$ (curve labelled ``C''), which for $N=2$ electrons in 1D has been shown\cite{GroKooGieSeiCohMorGor-JCTC-17} to approximate very accurately the exact HK functional at large $\lambda$.
The analogous of Eq.~\eqref{eq:taulambdaapprox} cannot be derived analytically, but we use for the asymptotics of $\tau(\lambda)$ at high couplings the dependence discussed in Sec~\ref{sec:ineq} and confirmed in eq~\eqref{eq:taulambdaapprox}, namely
\begin{equation}\label{eq:taususpect}
\tau(\lambda)\sim a\sqrt{\frac{1}{\lambda}},\quad\lambda\to\infty,
\end{equation} 
and we optimize $a$ in order to match the expansion of the HK functional. We get $a\approx 0.27$ for $v_{ee}^{\mathrm{reg}}(x)$ and a very similar value, $a\approx 0.32$, for  $v_{ee}^{\mathrm{exp}}(x)$. The curve labelled ``B'' shows the corresponding $G_{\lambda}^{\tau(\lambda)}[\dens]$ when we set $\tau(\lambda)$ equal to eq~\eqref{eq:taususpect}. In the curve labelled ``A'' we have simply set $\tau=\lambda^{-1}$, which was the small-$\lambda$ expansion found for the harmonic interaction case. We also show in the same figure the left-hand side of the inequality~\eqref{eq:ineqsuSCE} when we set $\tau(\lambda)=\frac{\pi}{2\lambda}$, which was found in the inequality \eqref{eq:SobolevInequality}, curve labelled ``D''. As it should, this curve stays above the value of $V_{ee}^{\rm SCE}[\dens]$ (horizontal line, labelled ``F''), but, contrary to the harmonic case of fig~\ref{fig:companalitico}, this time this curve does not stay below the $\lambda$-dependent HK functional. We also show the right-hand side of the inequality \eqref{eq:SobolevInequality} (curve labelled ``E''), which, again is found to be a very loose lower bound.

\section{Conclusions and Outlook}\label{sec:conc}
In this paper, we introduced and studied structural properties of a new class of density functionals based on the entropic-regularization of the SCE functional. Although the entropic regularization of the OT-SCE problem has been previously used as a numerical tool to compute the SCE energy via the Sinkhorn algorithm, here we have investigated whether it could also provide a route to build and study approximations of the Hohenberg-Kohn functional at large coupling constant $\lambda$. We  have first focused on the link between the (classical) entropy with fixed marginals used here, the quantum kinetic energy, and the Kullback–Leibler divergence, with links to the seminal work of Sears, Parr and Dinur~\cite{SeaParDin-IJC-80}, and with other recent works in the same spirit.\cite{Del-IJQC-15,Nag-CPL-13,NagLiu-PRA-08,SagLagGue-CPL-11,GonFerDeh-PRL-03,MolEsqAngAntDeh-JMC-12,BorJaqTorVanGee-PCCP-09,WelMerBalHol-PCCP-14,GotMau-PRL-05}

We have performed a very preliminary investigation on whether the minimizing wave function of the regularized SCE entropic problem, which has an explicit form, could be used to estimate the kinetic energy. A more extensive investigation is needed, in order to assess whether it is possible to find an approximate general relation between $\tau$ and $\lambda$, at least for large $\lambda$. We conjectured here, and we have numerical evidence in very simple cases, that when $\lambda\to\infty$ it holds $\tau\sim a\lambda^{-1/2}$, with $a$ probably a density-dependent constant.

We should remark that, from a computational viewpoint,  a challenging problem is to face the very unfavorable scaling with respect to the number of electrons (marginals) $N$ of the Sinkhorn algorithm when solving the entropic-SCE problem \cite{LinHoCutJor-ARXIV-19}. This implies that in order to provide functionals for routine applications, we might need to construct approximations inspired to the mathematical form of eq~\eqref{eq:Tentrexplicit}, similar to what has been done for the leading SCE term.\cite{WagGor-PRA-14,BahZhoErn-JCP-16,VucGor-JPCL-17,Vuc-JCTC-19,VucGou-JCP-19} To this purpose, it will be essential to further study properties of $u^\tau$ at small $\tau$, also comparing and testing it as a candidate for the Hartree-exchange-correlation potential.

\section{Acknowledgements}
Financial support was provided by the H2020/MSCA-IF ``OTmeetsDFT'' [grant 795942] and the European Research Council under H2020/ERC Consolidator Grant ``corr-DFT'' [grant 648932]. We also thank S. Di Marino and L. Nenna for fruitful discussions. All the authors contributed equally to this work.

\appendix
\section{Dual formulation}
While for the technical details and the rigorous proof we refer the reader to ref~\cite{GerKauRaj-ARXIV-19}, here we just want to give a rough idea of why the optimal $\gamma$ takes the form \eqref{eq:aj}.
Consider $\gamma \in \MM(\R^{3N})$, $u\in C_0(\R^3)$ and define 
\begin{equation}
\Epot(x_1,\dots,x_N) = V_{ee}(x_1,\dots,x_N) - \sum^N_{i=1}u(x_i). 
\end{equation}
Then,
\begin{align*}
E^{\tau}[\gamma] &= \int_{\R^{3N}}V_{ee}\gamma \dxN + \tau\int_{\R^{3N}}\gamma \log\gamma \dxN \\
&= \int_{\R^{3N}}\Epot(x_1,\dots,x_N)\gamma \dxN + \sum^N_{i=1}\int_{\R^{3N}} u(x_i)\gamma \dxN  \\&+ \tau\int_{\R^{3N}}\gamma \log\gamma\dxN \\ 
&= \int_{\R^{3}} u(x)\dens(x)\mathrm{d}x+ \int_{\R^{3N}}(\Epot(x_1,\dots,x_N)+\tau\log\gamma(x_1,\dots,x_N))\gamma \dxN \\
&\geq \int_{\R^{3}} u(x)\dens(x)\mathrm{d}x - \tau\int_{\R^{3N}} \exp\left(-{\frac{\Epot(x_1,\dots,x_N)}{\tau}}\right)\dxN + \tau,
\end{align*}
where we used $t s + \tau t \log t - \tau t \geq -\tau e^{-s/\tau}$, with equality if $t = e^{-s/\tau}$. Therefore,
\[
F_{\rm entr}^{\tau}[\dens] = \min_{\gamma\to\dens}E^{\tau}[\gamma] = \sup_{u^{\tau} \in C_0(\R^3)} \left\lbrace \int u^{\tau}(x)\dens(x)\mathrm{d}x  - \tau\int_{\R^{3N}}e^{-\frac{E^\tau_{pot}(x_1,x_2)}{\tau}}\dxN \right\rbrace +\tau.
\]
Notice that by writing $a^\tau(x_j) = e^{\frac{u^\tau(x_j)}{\tau}}$, one can associate the entropic weights with the entropic potentials $u^\tau(x_j)$.

%\bibliography{bib_clean}

%\iffalse

\providecommand{\latin}[1]{#1}
\providecommand*\mcitethebibliography{\thebibliography}
\csname @ifundefined\endcsname{endmcitethebibliography}
  {\let\endmcitethebibliography\endthebibliography}{}

\end{document}